\documentclass[aps,prb,twocolumn,showpacs,groupedaddress]{revtex4-1}

\usepackage{graphicx}
\usepackage{dcolumn}
\usepackage{bm}
\usepackage{amsmath}
\usepackage{natbib}
\usepackage{lineno}
 \usepackage{array}

\begin{document}

\preprint{APS/123-QED}

\title{Effect of picosecond strain pulses on thin layers of the ferromagnetic semiconductor (Ga,Mn)(As,P)}

\author{L. Thevenard$^{1}$\email[e-mail: ]{thevenard@insp.jussieu.fr}, E. Peronne$^{1}$, C. Gourdon$^{1}$, C. Testelin$^{1}$, M. Cubukcu$^{1}$, E. Charron$^{1}$, S. Vincent$^{1}$, A. Lema\^{\i}tre$^2$, and B. Perrin$^{1}$}

 
\affiliation{
$^1$ Institut des Nanosciences de Paris (CNRS), Universit\'{e} Pierre et Marie Curie, UMR7588, 140 rue de Lourmel, Paris, F-75015 France\\
$^2$ Laboratoire de Photonique et Nanostructures, CNRS, UPR 20, Route de Nozay, Marcoussis, F-91460 France}

\date{\today}

\label{sec:Abstract}

\begin{abstract}

The effect of picosecond acoustic strain pulses (ps-ASP) on a thin layer of (Ga,Mn)As co-doped with phosphorus was probed using  magneto-optical Kerr effect (MOKE). A transient MOKE signal followed by low amplitude oscillations was evidenced, with a strong dependence on applied magnetic field, temperature and ps-ASP amplitude. Careful interferometric measurement of the layer's  thickness variation induced by the ps-ASP allowed us to model very accurately the resulting signal, and interpret it as the strain modulated reflectivity  (differing for $\sigma _{\pm}$ probe polarizations), independently from dynamic magnetization effects.

\end{abstract}


\maketitle

\section{INTRODUCTION}

After a decade of research on the dilute magnetic semiconductor (DMS)  GaMnAs leading to record Curie temperatures ($T_{C}$) of about 200~K \cite{Novak2008}, one of the current challenges is to manipulate the magnetization on very short time-scales, as is routinely done in ferromagnetic metals\cite{Schumacher03,Hohage08} in view of ultra-fast switching in magnetic storage devices. A number of groups have used rapid thermal or carrier concentration modulation with ultra-fast light pulses to that end\cite{Kimel2004,Rozkotov'a2008,Wang2007a}. In this work, we investigate the possibility of rapid modulation of the magnetism of GaMnAs using picosecond acoustic strain pulses (ps-ASP). The specific coupling between localized Manganese spins and delocalized carriers leads  to an acute dependence of the magnetic anisotropy energy and  $T_{C}$ on the geometry and filling of the valence band, mainly governed by temperature, carrier and Mn concentrations, or strain \cite{Dietl2001,thevenard07}. The magnetic anisotropy of GaMnAs has thus already been tuned quasi-statically through strain by growth engineering\cite{Thevenard2006} or  by fitting strain actuators onto the layer\cite{Bihler2008, Masmanidis2005}. 

For ultra-fast strain modifications however, acoustic waves are a more relevant tool, piezo-electric transducers being limited to frequencies of about 1~GHz.  Predicted half a century ago by Kittel \cite{Kittel1958}, extensive work in ferromagnetic metals\cite{bommel59,Melnikov2003} has  evidenced the interactions between lattice vibrations  and spin-waves, with the emitted hypersonic power reaching a maximum at the ferromagnetic resonance field for which the magnetization precesses uniformly ($\omega _{phonon}$=$\omega _{precession}$). In  DMS, propagation of acoustic waves has been reported \cite{Qi2010,Wang2005b}, and only one very recent study is proposing magnetization-dependent effects under acoustic strain pulse excitation\cite{scherbakov10cm}. Fundamental questions linked to the carrier-mediated nature of the ferromagnetism in these materials thus remain open, such as the mechanism, dynamics and efficiency of the coupling of lattice vibrations  to spin waves via carriers.

In this work, we report the effect of a ps-ASP on GaMnAs codoped with phosphorus. Strain and magnetization were  studied thanks to the coupling of a Sagnac interferometer to a time resolved magneto-optical polar Kerr effect (TR-MOKE) set-up. Careful ferromagnetic resonance (FMR)  characterization allowed us to rule out the  observed frequencies as Mn spin precession. The detected TR-MOKE signal was simulated using a multi-layer thin  film reflectivity model which showed that the field, temperature and magnetization dependence of the data could be fully explained by the measured ps-ASP shape and the Zeeman splitting dependence of the optical index.

\section{SAMPLE and METHODS}

GaMnAs was alloyed with phosphorus, adjusting its concentration to obtain both a weak magnetic anisotropy \cite{lemaitre08,Cubukcu2010} and a weak strain. A $d$=50nm layer of (Ga$_{0.93}$Mn$_{0.07}$)(As$_{0.93}$P$_{0.07}$) was grown on a double-polished GaAs substrate, the dopant concentrations being determined as in Ref.\onlinecite{lemaitre08}. The layer was annealed for 1h at 250$^{\circ}$C. The perpendicular lattice parameter $a_{\bot}$ was determined by high-resolution X-Ray diffraction, evidencing a  weakly tensile layer with a strained lattice mismatch -2660~ppm, giving $\epsilon_{zz}=(a_{\bot}-a_{relaxed})/a_{relaxed}$=$-0.127\%$.  Finally, 30~nm of  Aluminium  were deposited on the substrate side to be used as acoustic transducer, and on half of the magnetic layer to be used as a a mirror for surface displacement measurements. Superconducting Quantum Interference Device measurements yielded $T_{C}$=116~K, and an effective Mn concentration of $x^{eff}_{Mn}=5\%$. The magnetic anisotropy constants   obtained by FMR evidenced a very weak in-plane easy axis. At 40~K, the uniaxial anisotropy field amounted to  $B^{0}_{2\bot}=2K_{2\bot}/M$=521~G, while the rest of the anisotropy coefficients were $B_{2//}$=154~G, $B_{4\bot}$=277~G and $B_{4//}$=128~G. Adjusting the phosphorus content for a weak strain yielded a soft magnetic layer where magnetization can be more easily manipulated.

The sample was placed in a variable temperature He cryostat. The average temperature increase due to the average pump fluence was measured on a Al/GaAs test sample and found to be of $\Delta T$=17~K  at  $T_{0}$=10~K, and $\Delta T$=7~K at $T_{0}$=81~K, for \textit{I}=370~$\mu$Jcm$^{-2}$. Working at  high temperatures (50-80~K) therefore limited the temperature increase with fluence. In this paper, only effective sample temperatures are given.

\begin{figure}
	\centering
		\includegraphics[width=0.5\textwidth]{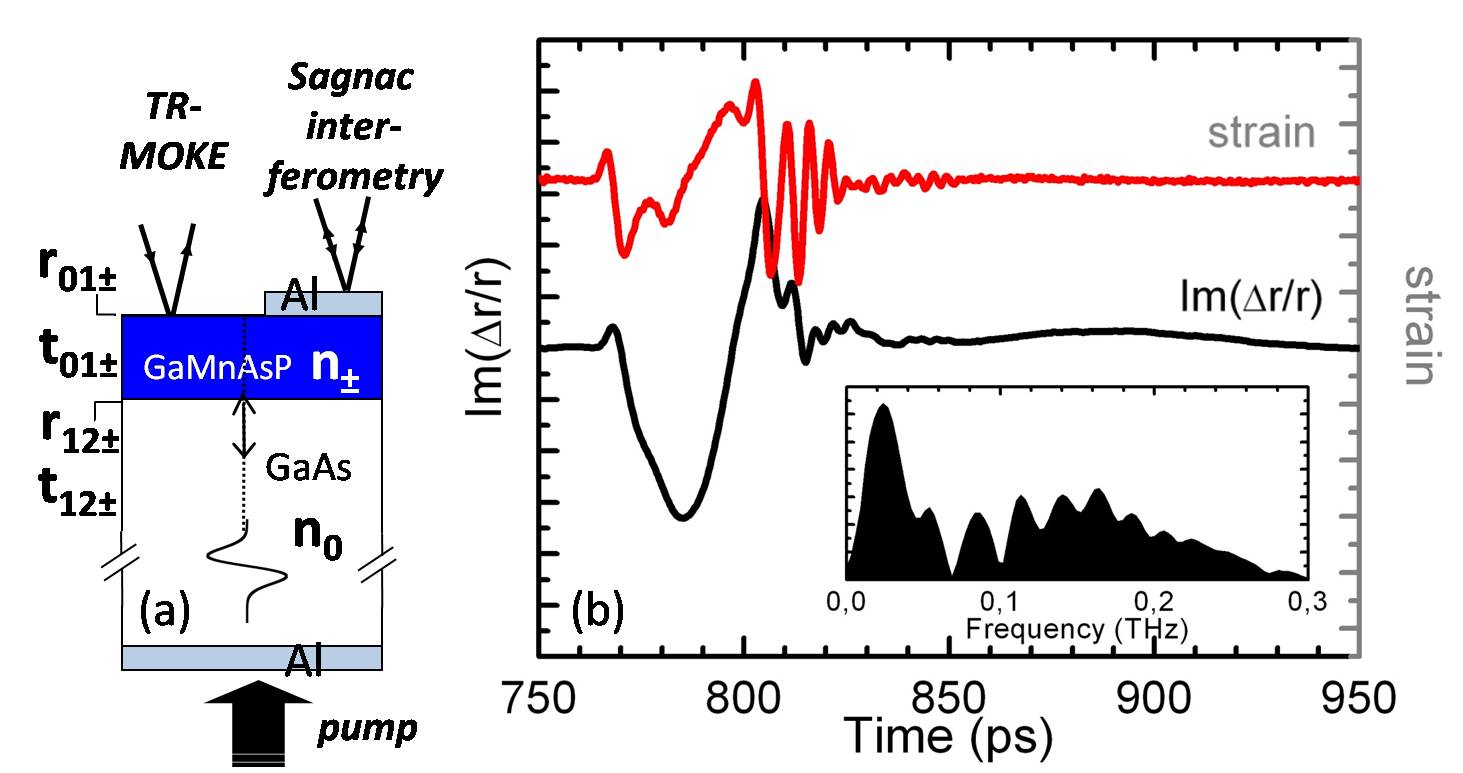}
	\caption{(a) Schematics of the set-up with the probe beam on Al for the acoustics characterization, or directly on GaMnAsP for MOKE measurements. Definitions of the reflection, and transmission coefficients. (b) Interferometry characterization of the acoustic strain pulse ($T$=58~K, $\lambda_{probe}$=780~nm, $I$=280~$\mu$Jcm$^{-2}$): surface displacement Im($\Delta r /r$) and corresponding strain incident  on the GaMnAsP layer. Inset: strain spectrum. }
	\label{acous}
\end{figure}

\section{ACOUSTIC RESULTS}

The ps-ASP are generated using a standard picosecond acoustic scheme based on ultrafast transduction in a metallic thin film\cite{peronne06}. A first Ti:Sapphire laser delivering 130~fs pulses at $f_{pump}=$~80~MHz (780~nm) generates longitudinal strain wave packets through the back-side transducer. A second Ti:Sa laser delivers  80~fs pulses at $f_{probe}=f_{pump} + \Delta f $. They were used to detect the acoustic strain pulse  after propagation through the substrate or the magnetization change induced by the strain pulse on the front-side GaMnAsP layer (Fig. \ref{acous}a). This asynchronous optical sampling (ASOPS) scheme based on the so-called stroboscopic effect  has recently been implemented in  a picosecond acoustic experiment \cite{klatt09}. This alternative technique is much less sensitive to the mechanical issues encountered in traditional delay line set-ups, and it allows shorter acquisition times and independent tuning of pump/probe wavelengths. The pump-probe delay range is given by $1/f_{pump}$ and the step size by $\Delta f/[f_{pump}(f_{pump}+\Delta f)]$. The time resolution, which is mainly limited by the jitter of the lasers, was estimated around 0.5 to 2~ps  for a 0-3~ns time window (i.e. experimental bandwidth of about 250 GHz). The main features and  details of this experimental set-up will be published elsewhere.

Fig. \ref{acous}b (black line) displays the phase change of the reflectivity  $\Delta r / r$ measured by a Sagnac interferometer on the Al deposited part of the sample. The probe beam was linearly polarized, with a fluence of 10~$\mu$Jcm$^{-2}$ and focused to a spot size comparable to the pump spot size. Previous work\cite{peronne06} has shown that the phase change is mainly due to surface displacement. Therefore the incoming strain can be retrieved by taking into account acoustic reflection on the Al layer. The calculated strain pulse stressing the GaMnAsP layers is shown in Fig. \ref{acous}b (gray line) and displays the typical features observed in non-linear ps acoustics: two acoustic solitons in front and an oscillating tail. The pump spot size allowed us to neglect the diffraction effect during the propagation.  By fitting the incoming ps-ASP with a non-linear propagation equation\cite{peronne06}, the soliton amplitude is found to be of the order of 8.4.10$^{-4}$ for pump fluence \textit{I}=280~$\mu$Jcm$^{-2}$ at 58~K. The maximum accessible strain is therefore slightly lower than the static piezo-induced uniaxial strain of Bihler et al.\cite{Bihler2008}, but occurs on a much shorter timescale of about 2-10~ps. The strain spectrum is shown as an inset of Fig. \ref{acous}b. Its shape is typically broad (over 300 GHz) with some modulation due to non-linear effects. The higher frequencies  are cut off mainly by the time jitter and pump-probe spatial averaging. Note that the most intense part of the spectrum is located around 20 GHz, reasonably close to the  fundamental and excited frequencies of most magnetic systems. Frequencies below 50~GHz are barely affected by sound absorption through 360~$\mu$m thick GaAs up to 100 K. Moreover we checked that the ps-ASP shape and amplitude remain unchanged under a perpendicular-to-plane magnetic field of $B$=$\mu_{0} H$=$\pm$50~G. Hence the acoustic pulse shown in Fig.  \ref{acous}b can safely be used to understand the results obtained for various magnetic fields and sample temperatures and discussed in the following.

\section{TR-MOKE RESULTS}

\begin{figure}
	\centering
		\includegraphics[width=0.5\textwidth]{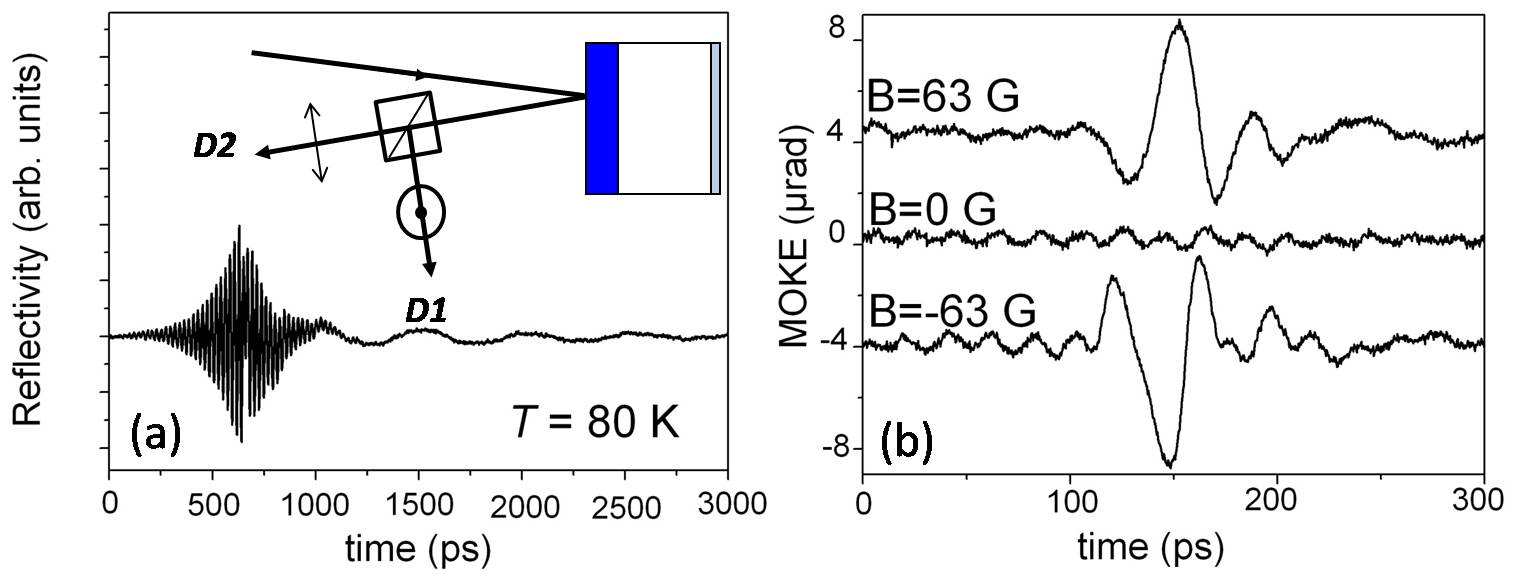}
	\caption{(a) Schematics of the optical bridge placed on the path of the reflected beam, and reflectivity seen by one of the diodes ($B$=63~G, $I$=280~$\mu$Jcm$^{-2}$, $\lambda_{probe}$=780~nm, $T$=80~K). (b) Transient Kerr signal under opposite applied fields ($I$=280~$\mu $Jcm$^{-2}$, $\lambda_{probe}$=705~nm, $T$=58~K). Curves were spaced equally for clarity, and $t$=0 set arbitrarily.}
	\label{fig:Kerr+-63Oe}
\end{figure}

\label{sec:EffetDUnChampMagnétique}

\begin{figure}
	\centering
		\includegraphics[width=0.4\textwidth]{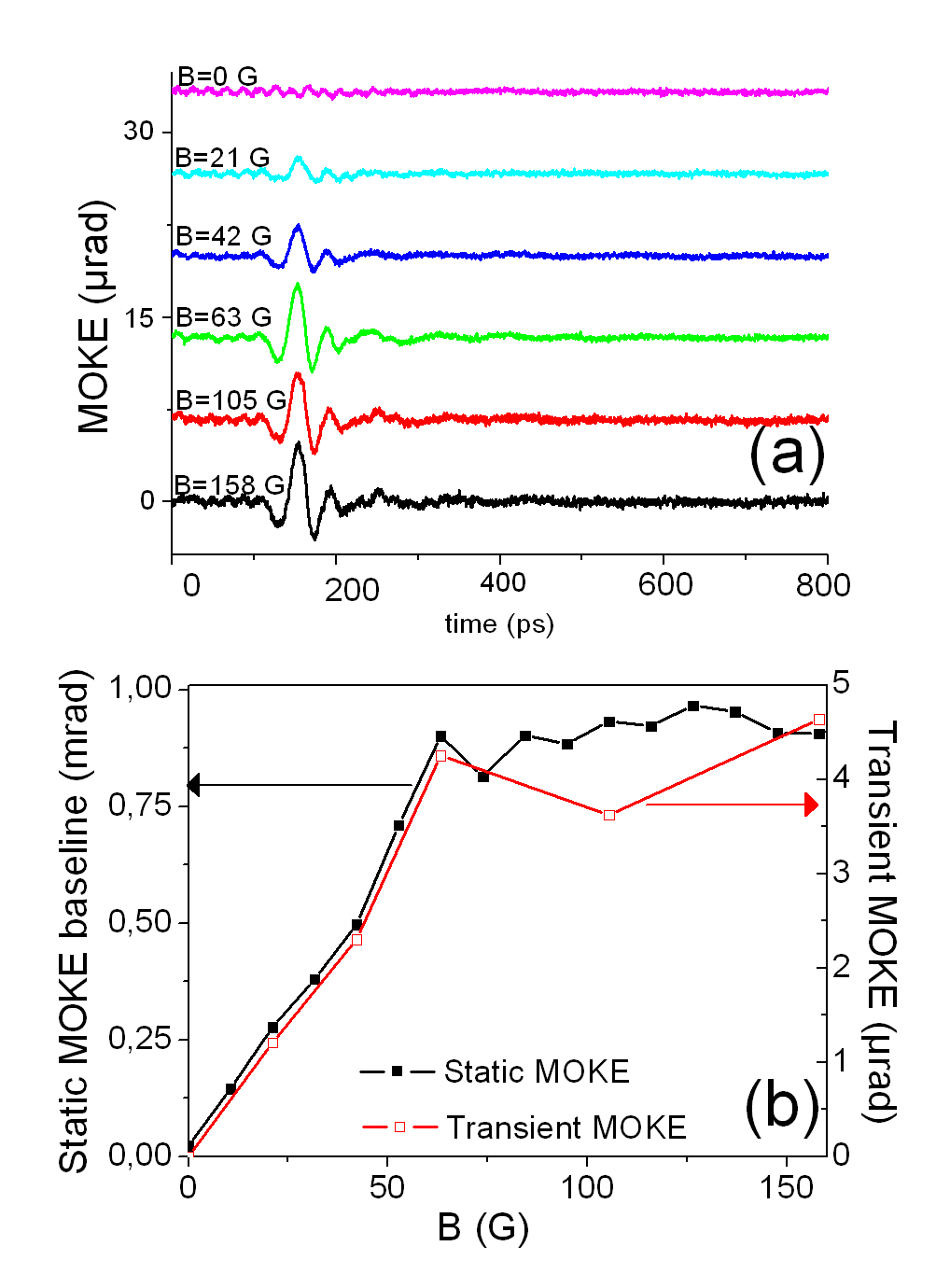}
	\caption{(color on-line) (a) Transient MOKE signal with varying $B_{\bot}$. ($T$=58~K, $\textit{I}$=280~$\mu$Jcm$^{-2}$, $\lambda$=705~nm). Curves were spaced equally for clarity, and $t$=0 set arbitrarily. (b) Transient and static MOKE versus applied field.}
	\label{fig:Champs-tous}
\end{figure}

The TR-MOKE signal was  recorded  on the Al-free part of the GaMnAsP layer. The rotation of the linear probe polarization was detected after reflection by a balanced optical bridge, after passing through a Glan laser polarizer. The reflectivity seen by each diode is of the type shown in Fig. \ref{fig:Kerr+-63Oe}a, and corresponds to interferences  between the probe beam and the one reflected off the ps-ASP within the GaMnAsP layer, giving $\approx$45~GHz Brillouin oscillations. The resulting Kerr signal is computed from the difference of the diode signals, with the Kerr rotation angle $\Theta_{K}$   proportional to the perpendicular component of the magnetization $M_{\bot}$. Under no applied field, the TR-MOKE signal was constant save for small amplitude oscillations, whereas when the field was switched on, a transient signal appeared (Fig. \ref{fig:Kerr+-63Oe}b). Its shape changed sign upon reversal of the field, with a slightly smaller amplitude and a 5~ps phase shift. 
For  $\textit{I}$=280~$\mu$Jcm$^{-2}$, its amplitude increased steadily with applied field, and saturated at $B_{sat}$=63~G (resp. 37~G) for $T$=58~K (resp. $T$=83~K)  as shown in Fig. \ref{fig:Champs-tous}. These correspond well to the saturation fields determined from $\Theta_{K}$(B) curves obtained from the static Kerr baseline value: 53~G for $T$=58~K and 37~G for $T$=83~K   (Fig. \ref{fig:Champs-tous}b). At $T$=130~K, the $B$=0, and $B$=$\pm$63~G traces were identical, indicating the disappearance of this phenomenon above the Curie temperature. Finally, no transient Kerr signal could be observed at all when an in-plane field of up to 420~G was applied along the [110] axis of the layer.

\begin{figure}
	\centering
		\includegraphics[width=0.4\textwidth]{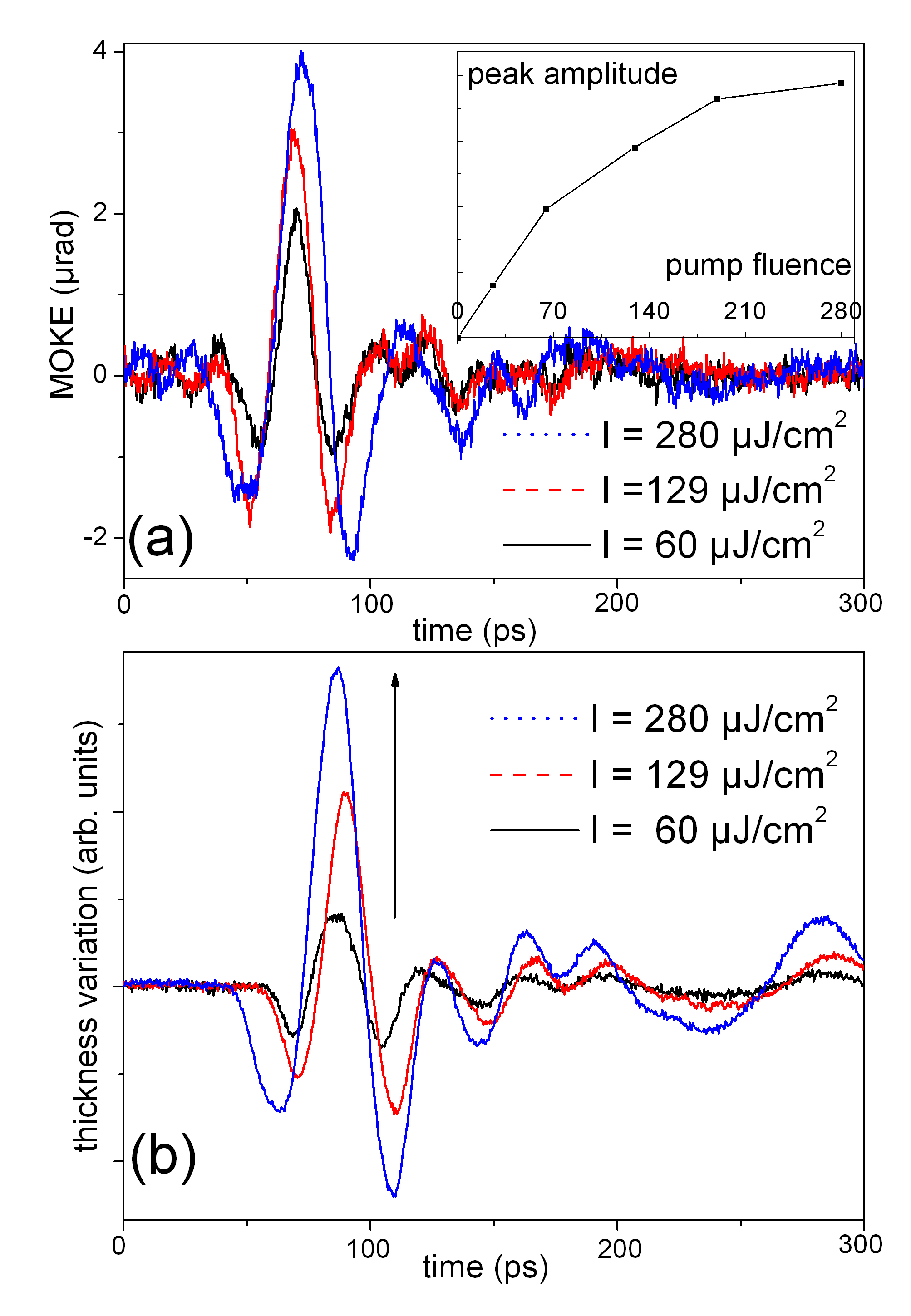}
\caption{(color on-line) Varying pump fluences, $\textit{I}$=60-280~$\mu$Jcm$^{-2}$. $T$=54$\pm$4~K, $\lambda_{probe}$=705~nm, $t$=0 set arbitrarily. (a) Transient MOKE signal at $B$=63~G. Inset: peak amplitude versus fluence. (b) GaMnAsP layer thickness variation estimated from the Sagnac interferometry data.} 
	\label{fig:2phi}
\end{figure}

The perpendicular magnetic field was then set to a fixed value of $B$=63~G, and the pump fluence varied between $\textit{I}$=26-280~$\mu$Jcm$^{-2}$ to progressively increase the amplitude of the ps-ASP. The main transient Kerr signal then broadens, arriving earlier with increasing pump fluence (Fig. \ref{fig:2phi}a), with frequencies around 20-30~GHz. Small amplitude 10~GHz oscillations appear after the main peak. Note that the 2~GHz oscillations observed in the reflectivity at $T$=80~K (Fig. \ref{fig:Kerr+-63Oe}a), $\lambda$=780~nm were not reproduced regularly, or observed in the Kerr data at $T$=50~K or 80~K, $\lambda$=705~nm.  The peak amplitude increases monotonously with pump fluence (Fig. \ref{fig:2phi}a, inset). For comparison, we plot the thickness variation of the GaMnAsP layer at varying fluences, obtained by integrating the strain over its thickness (Fig. \ref{fig:2phi}b). The MOKE and thickness variation traces clearly present very similar features at all fluences. A 10~GHz component also appeared in the interferometry data at $\lambda_{probe}$=705~nm. Its origin is to this day unclear, but was found non-magnetic since it has also been observed on non-intentionally doped GaAs substrates\cite{osc}.

\section{DISCUSSION}

The possible origins of the transient MOKE signal will now be discussed. MOKE corresponds to a phase difference between reflectivities of   $\sigma^{+}$ and $\sigma^{-}$ circular polarizations off a magnetic material. For small rotation angles and weak absorption, the Kerr rotation angle can be expressed as $\Theta_{K}\approx\frac{\kappa_{+}-\kappa_{-}}{\tilde{n}^{2}-1}$ with the optical index $n_{\pm}$=$\tilde{n}_{\pm}+i\kappa_{\pm}$, $\tilde{n}^{2}\approx \tilde{n}_{+}\tilde{n}_{-}$, and $\Theta_{K}\propto M_{\bot}$. A first interpretation is to assume the  strain modification induced by the ps-ASP strong enough to alter the uniaxial magnetic anisotropy field $B_{2\bot}$ of the layer, and trigger a magnetization precession. It has been shown that the biaxial strain has very little influence on the other magnetic anisotropy fields\cite{Cubukcu2010}.  Given the typical strain modifications determined above,  variations of up to $\Delta B_{2\bot}\approx$~250~G (half the initial anisotropy) can be expected from both a phenomenological $B_{2\bot}$($\epsilon_{zz}$)  relationship\cite{Glunk2009} or magneto-strictive coefficients \cite{Masmanidis2005}. The Landau-Lifshitz-Gilbert equation can then be solved numerically with $B_{\bot}<B_{sat}$, the anisotropy fields at $T$=40~K determined from FMR and   $B_{2\bot}(t)=B^{0}_{2\bot} + \Delta B_{2\bot}(t)$, the latter having the same profile as the propagating strain pulse. The magnetization is predicted to start to precessing uniformly at $t$=0 around $\vec{M}_{0}$  at the  zero-momentum magnon frequency  $f_{R}$=213~MHz with $\Delta M_{\bot}/M_{\bot}$=5.10$^{-4}$, slightly above our experimental sensitivity. This frequency is very low due to the weak anisotropy and high temperature; it has not come out in our data. The observed frequencies did not disappear at saturation, and therefore do not correspond to higher-order spin-wave modes. Various inhomogeneities may have prevented us from observing precession. A multi-domain configuration was excluded  by cooling the sample down under a 100~G in-plane magnetic field. Low frequency noise may have prevented us from seeing the low frequency oscillations. Finally, the reduced power of the low frequency ($<$~1~GHz) part of the strain spectrum may have lead to an inefficient excitation of the manganese spins.

 As in metals\cite{Koopmans2000}, transient Kerr rotation can also  be obtained at constant magnetization $\vec{M}$ from spurious magneto-optical effects. We have thus excluded  Faraday rotation (circular birefringence) in the  substrate, unlikely under the small applied fields,  and magnetic linear dichroism\cite{Kimel2005}, excluded by rotating the impinging linear polarization of the probe along the layer's high symmetry axes. Finally, the generation of hot photo-carriers in the GaMnAsP layer is prevented by pumping in the highly  absorbing Al back-layer, and the linear polarization of the probe beam insures no angular momentum is being injected. The main effect of the pump is to modify $a_{\bot}$.

 A thin film multi-layer reflectivity model was then implemented  using the transfer matrix method\cite{perrin96}, in order to introduce the presence of the GaMnAsP/GaAs interface, the reflectivity difference for $\sigma_{\pm}$ polarizations, as well as the finite width of the ps-ASP with respect to the layer thickness. The reflectivities for $\sigma_{\pm}$ polarizations are given by $r_{\pm}=r_{0\pm}+\Delta r_{\pm}$, with $\Delta r_{\pm}$ the sum of 4 terms as follows: \\
 
 \begin{equation}
   \Delta r_{\pm} =   S_{\pm}+A_{\pm}+B_{\pm}+C_{\pm} 
  \end{equation}
  
  \begin{eqnarray}
  S_{\pm} & = & ik2u(0,t)r_{0\pm}, \\
  A_{\pm} & = & ik(2n^{2}_{\pm}+2n_{\pm}\frac{\partial n_{\pm}}{\partial \eta})a_{\pm}b_{\pm}\int^{d}_{0}\eta(z,t)dz, \\
    B_{\pm} & =&  \\
    & &   ik2n_{\pm}\frac{\partial n_{\pm}}{\partial \eta} \int^{d}_{0}(a_{\pm}^{2} e^{2ikzn_{\pm}} + b_{\pm}^{2} e^{-2ikzn_{\pm}})\eta(z,t)dz, \nonumber \\
   C_{\pm} & = & ik2n_{0}\frac{\partial n_{0}}{\partial \eta}\int^{\infty}_{d} a^{2}_{s \pm}e^{2ik(z-d)n_{0}}\eta(z-d,t)dz
    \end{eqnarray} \\

Where the coefficients are given by:
 
       \begin{eqnarray*}
   a_{\pm} & = & \frac{t_{01\pm}e^{-i\alpha_{\pm}d}}{e^{-i\alpha_{\pm}d}+r_{01 \pm}r_{12 \pm}e^{i\alpha_{\pm}d}}, \\
  b_{\pm}  & = & \frac{r_{12 \pm}t_{01 \pm}e^{i\alpha_{\pm}d}}{r_{01 \pm}e^{-i\alpha_{\pm}d}+r_{12 \pm}e^{i\alpha_{\pm}d}}, \\
  a_{S \pm}& = & \frac{t_{01 \pm}t_{12 \pm}}{e^{-i\alpha_{\pm} d}+r_{01 \pm}r_{12 \pm}e^{i\alpha_{\pm} d}}, \\
   r_{0 \pm}& = &  r_{01 \pm} + t^2_{01 \pm}r_{12 \pm}e^{2i\alpha_{\pm}d}
    \end{eqnarray*} \\

The transmission  and reflection coefficients  at the air/GaMnAsP interface are $t_{01\pm}$, $r_{01\pm}$ , and  $t_{12\pm}$, $r_{12\pm}$ at the GaMnAsP/GaAs  interface (Fig. \ref{acous}a). The optical index in GaAs is $n_{0}$, and $n_{\pm}$ within the GaMnAsP layer, with mean values obtained by  room-temperature ellipsometry measurements. The absorption of both layers was defined by the imaginary part of the index:  $\alpha_{\pm}=2k_{0}\kappa_{\pm}$ and  $\alpha_{0}$. The beam was taken normally incident to the layer with $k$  the vacuum wave-vector.  The propagating strain shape $\eta(z,t)$ was calculated as explained above from the surface displacement data $u(0,t)$ over a 400~ps window, with $t$=0 taken approximately as the arrival time of the strain wave upon the surface ($I$=280~$\mu$Jcm$^{-2}$, $\lambda$=705~nm, $T$=58~K, $B$=63~G). The photo-elastic constants of GaMnAsP and GaAs were  taken equal throughout the layer, with 
 $\frac{\partial n_{+}}{\partial \eta}$=$\frac{\partial n_{-}}{\partial \eta}$=$\frac{\partial n_{0}}{\partial \eta}$. 

 Term $S_{\pm}$ is the usual reflectivity change associated with the displacement of the surface\cite{perrin96}. Term $A_{\pm}$ is proportional to the thickness variation of the layer. Term $B_{\pm}$ (resp. $C_{\pm}$) corresponds to the afore-described Brillouin oscillations within the GaMnAsP (resp. GaAs layer). The observed signal is then computed as :
 
 \begin{equation*}
 \begin{split}
 \Delta I & =|(1-\epsilon)(r_{+}+r_{-})-i(1+\epsilon)(r_{+}-r_{-})|^{2}\\ 
 & -|(1+\epsilon)(r_{+}+r_{-})+i(1-\epsilon)(r_{+}-r_{-})|^{2}
\end{split}
  \end{equation*}
 
   A possible unbalance $\epsilon$ of the optical bridge is included, taking $\epsilon$=0 when both diodes see the exact same signal in the absence of Kerr effect ($T>T_{C}$, $M_{\bot}$=0  or $M$ in the plane of the sample). The actual Kerr angle can then be estimated by $\Theta_{K}\approx
	\frac{\Delta I}{4I}$. 
	
\begin{figure}
	\centering
		\includegraphics[width=0.50\textwidth]{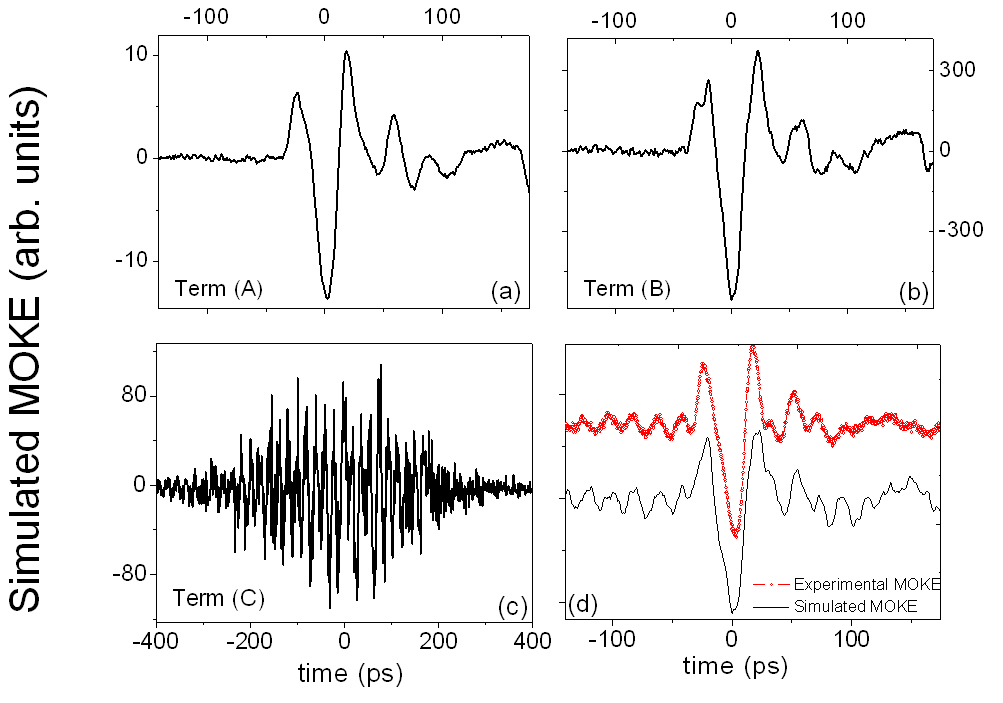}
	\caption{(color on-line)  a), (b), (c) Normalized simulated MOKE signal using a multi-layer reflectivity model. See within text for
	 description of the different terms. (d) Comparison of the full simulated MOKE signal with the experimental data ($\textit{I}$=280~$\mu$Jcm$^{-2}$, $T$=58~K, $\lambda_{probe}$=705~nm, $B$=-63~G).}
	\label{fig:Simul}
\end{figure}

	In Fig. \ref{fig:Kerr+-63Oe}a, the $B_{\bot}$=0 curve shows   oscillations of small amplitude ($\approx$1~$\mu$rad), although the magnetization is expected to be in plane. Taking $n_{+}$=$n_{-}$ in the expression of $\Delta I$, a bridge unbalance of merely $\epsilon$=2$\%$ is enough to simulate oscillations of this amplitude. They correspond to a slight phase difference between reflectivities on diodes 1 and 2 (see Fig. \ref{fig:Kerr+-63Oe}a). Once this value of $\epsilon$ determined, the difference $\Delta n$=$n_{+}-n_{-}$ was adjusted to yield the experimentally observed static Kerr angle of $\approx$1~mrad and resulted in $\Delta n$=0.05-0.0126$i$ ($\Delta \kappa$ estimated beforehand from independent magnetic circular dichroism measurements). Finally, the  terms in the signal $I$ corresponding to expressions (3), (4) and (5)  were normalized to ease comparison and plotted separately as terms (A), (B) and (C) in Fig. \ref{fig:Simul}a,b,c. Term (S) is negligeable since the corresponding measured signal will be the product of the squared surface displacement by the static Kerr angle. Term (A) follows exactly the thickness variation profile, with a weak relative amplitude reflecting the small optical index mismatch between the thin layer and its substrate. Term (C) has the typical shape of Brillouin oscillations, where the ps-ASP is seen to come towards the surface, and then away,  quite like the single-diode reflectivity data shown in Fig. \ref{fig:Kerr+-63Oe}a. The larger term is (B), with less than 1$\%$ of the signal coming from the $b_{\pm}^{2}$ return term from the GaMnAsP/GaAs interface. The total simulated TR-MOKE signal is then plotted against the experimental data in Fig. \ref{fig:Simul}d.

The simulation is very reminiscent of the data, with a broad transient MOKE peak preceded by small amplitude $\approx$45~GHz  Brillouin oscillations. The MOKE features following this large peak follow closely that of the simulated trace, relying only on reflectivity  calculations, independently of any precessional phenomenon. The model evidences that the signal comes mainly from the modulation of the reflectivity by the strain pulse, which affects slightly differently $\sigma _{+}$ and $\sigma _{-}$ probe polarizations, and is therefore picked up by the optical bridge. The simulated MOKE  amplitude  increases with the amplitude of $\eta (z,t)$ used in the model, experimentally dependent on the pump power. This explains the monotonous increase of the peak amplitude found with pump fluence (Fig. \ref{fig:2phi}a). The amplitude of the simulated MOKE peak also varies with the injected value  of  $\Delta n$, which can be varied by the sample temperature, or the applied magnetic field to modify the Zeeman splitting. This agrees well with the saturation of the transient Kerr signal amplitude at $B_{sat}$ (Fig. \ref{fig:Champs-tous}b) for which the Zeeman splitting saturates. The modification of the transient TR-MOKE signal   upon magnetic field reversal (Fig. \ref{fig:Kerr+-63Oe}b) can also be well explained qualitatively by considering  the different $\sigma^{\pm}$ contributions of the light and heavy hole bands to the absorption (and therefore to $\kappa_{\pm}$). When the field is reversed, so are the positions of the Zeeman-split  light/heavy hole bands, but the strain-induced splitting remains of the same sign, yielding a slightly asymmetrical behavior of $\kappa_{+}-\kappa_{-}$ (and therefore $\Theta_{K}$) with field, which we observed experimentally (Fig. \ref{fig:Kerr+-63Oe}b). This is particularly true as temperature is increased, and the Zeeman splitting decreases. 

Lastly, the carrier magnetization (a mere 5-10$\%$ of the total magnetization\cite{Dietl2001}) can also be modified through an ultra-fast valence band modulation upon arrival of the ps-ASP upon the layer. This is expected to give a second order effect on the optical index, through a $\sigma^{\pm}$ dependance of the photo-elastic constant in GaMnAsP. A numerical estimation of $\partial n_{\pm}/\partial \eta$ is however beyond the scope of this paper.

\section{CONCLUSION}

To summarize, the magnetization of a GaMnAsP layer was probed following  excitation by  a picosecond acoustic strain pulse. Contrary to the more standard ultra-fast pump-probe experiments on similar systems, the back-side pump did not create any carriers, allowing us to study the sole effect of the strain modulation. Thanks to an independent determination of the anisotropy coefficients, and therefore of the expected precession frequency, we can assert that the different frequencies appearing in the data did not correspond to uniform precession induced by inverse magneto-striction. A multi-layer reflectivity model showed that the MOKE signal could in fact be explained by the the strain pulse's modulation of the reflectivity, and it could be reproduced very well from the time evolution of the strain pulse; with the amplitude reflecting the strength of the Zeeman splitting and of the ps-ASP amplitude. This work highlights the necessity of both precise magnetic characterization and excellent knowledge of the shape of the picosecond strain pulses impinging on the layer in this type of experiment, and illustrates the versatility of the new alloy GaMnAsP.

We wish to acknowledge L. Largeau and O. Mauguin for the X-Ray measurements,  M. Bernard for the cryogenic facilities and B. Gallas for the ellipsometry data. This work was in parts supported by R\'{e}gion Ile de France under contract IF07-800/R with C'Nano IdF.

\bibliographystyle{phjcp}

\begin{thebibliography}{10}

\bibitem{Novak2008}
{\sc V.~Nov\'{a}k}, {\sc K.~Olejn\'{\i}k}, {\sc J.~Wunderlich}, {\sc M.~Cukr},
  {\sc K.~V\'{y}born\'{y}}, {\sc A.~Rushforth}, {\sc K.~Edmonds}, {\sc
  R.~Campion}, {\sc B.~Gallagher}, {\sc J.~Sinova}, and {\sc T.~Jungwirth},
\newblock {\em Phys. Rev. Lett.} {\bf 101} (2008).

\bibitem{Schumacher03}
{\sc H.~Schumacher}, {\sc C.~Chappert}, {\sc P.~Crozat}, {\sc R.~Sousa}, {\sc
  P.~Freitas}, {\sc J.~Miltat}, {\sc J.~Fassbender}, and {\sc B.~Hillebrands},
\newblock {\em Phys. Rev. Lett.} {\bf 90}, 017201 (2003).

\bibitem{Hohage08}
{\sc P.~E. Hohage}, {\sc J.~Nannen}, {\sc S.~Halm}, {\sc G.~Bacher}, {\sc
  M.~Wahle}, {\sc S.~F. Fischer}, {\sc U.~Kunze}, {\sc D.~Reuter}, and {\sc
  A.~D. Wieck},
\newblock {\em Appl. Phys. Lett.} {\bf 92}, 241920 (2008).

\bibitem{Kimel2004}
{\sc A.~V. Kimel}, {\sc G.~V. Astakhov}, {\sc G.~M. Schott}, {\sc A.~Kirilyuk},
  {\sc D.~R. Yakovlev}, {\sc G.~Karczewski}, {\sc W.~Ossau}, {\sc G.~Schmidt},
  {\sc L.~W. Molenkamp}, and {\sc T.~Rasing},
\newblock {\em Phys. Rev. Lett.} {\bf 92}, 237203 (2004).

\bibitem{Rozkotov'a2008}
{\sc E.~Rozkotov\'{a}}, {\sc P.~N\v{e}mec}, {\sc P.~Horodysk\'{a}}, {\sc
  D.~Sprinzl}, {\sc F.~Troj\'{a}nek}, {\sc P.~Mal\'{y}}, {\sc V.~Nov\'{a}k},
  {\sc K.~Olejn\'{\i}k}, {\sc M.~Cukr}, and {\sc T.~Jungwirth},
\newblock {\em Appl. Phys. Lett.} {\bf 92}, 122507 (2008).

\bibitem{Wang2007a}
{\sc J.~Wang}, {\sc I.~Cotoros}, {\sc K.~Dani}, {\sc X.~Liu}, {\sc J.~Furdyna},
  and {\sc D.~Chemla},
\newblock {\em Phys. Rev. Lett.} {\bf 98}, 1 (2007).

\bibitem{Dietl2001}
{\sc T.~Dietl}, {\sc H.~Ohno}, and {\sc F.~Matsukura},
\newblock {\em Phys. Rev. B} {\bf 63}, 195205 (2001).

\bibitem{thevenard07}
{\sc L.~Thevenard}, {\sc L.~Largeau}, {\sc O.~Mauguin}, {\sc A.~Lema\^{\i}tre},
  {\sc K.~Khazen}, and {\sc H.~von Bardeleben},
\newblock {\em Phys. Rev. B} {\bf 75}, 1 (2007).

\bibitem{Thevenard2006}
{\sc L.~Thevenard}, {\sc L.~Largeau}, {\sc O.~Mauguin}, {\sc G.~Patriarche},
  {\sc A.~Lema\^{\i}tre}, {\sc N.~Vernier}, and {\sc J.~Ferr\'{e}},
\newblock {\em Phys. Rev. B} {\bf 73}, 195331 (2006).

\bibitem{Bihler2008}
{\sc C.~Bihler}, {\sc M.~Althammer}, {\sc A.~Brandlmaier}, {\sc
  S.~Gepr\"{a}gs}, {\sc M.~Weiler}, {\sc M.~Opel}, {\sc W.~Schoch}, {\sc
  W.~Limmer}, {\sc R.~Gross}, {\sc M.~S. Brandt}, and {\sc S.~T.~B.
  Goennenwein},
\newblock {\em Phys. Rev. B} {\bf
  78}, 45203 (2008).

\bibitem{Masmanidis2005}
{\sc S.~C. Masmanidis}, {\sc H.~X. Tang}, {\sc E.~B. Myers}, {\sc M.~Li}, {\sc
  K.~D. Greve}, {\sc G.~Vermeulen}, {\sc W.~Van\~{}Roy}, and {\sc M.~L.
  Roukes},
\newblock {\em Phys. Rev. Lett.} {\bf 95}, 187206 (2005).

\bibitem{Kittel1958}
{\sc C.~Kittel},
\newblock {\em Phys. Rev.} {\bf 110}, 836 (1958).

\bibitem{bommel59}
{\sc H.~B\"{o}mmel} and {\sc K.~Dransfeld},
\newblock {\em Phys. Rev. Lett.} {\bf 3}, 83 (1959).

\bibitem{Melnikov2003}
{\sc A.~Melnikov}, {\sc I.~Radu}, {\sc U.~Bovensiepen}, {\sc O.~Krupin}, {\sc
  K.~Starke}, {\sc E.~Matthias}, and {\sc M.~Wolf},
\newblock {\em Phys. Rev. Lett.} {\bf 91}, 227403 (2003).

\bibitem{Qi2010}
{\sc J.~Qi}, {\sc J.~A. Yan}, {\sc H.~Park}, {\sc A.~Steigerwald}, {\sc Y.~Xu},
  {\sc S.~N. Gilbert}, {\sc X.~Liu}, {\sc J.~K. Furdyna}, {\sc S.~T.
  Pantelides}, and {\sc N.~Tolk},
\newblock {\em Phys. Rev. B} {\bf 81}, 115208 (2010).

\bibitem{Wang2005b}
{\sc J.~Wang}, {\sc Y.~Hashimoto}, {\sc J.~Kono}, {\sc A.~Oiwa}, {\sc
  H.~Munekata}, {\sc G.~Sanders}, and {\sc C.~Stanton},
\newblock {\em Phys. Rev. B} {\bf 72}, 1 (2005).

\bibitem{scherbakov10cm}
{\sc C.~B. {A.V. Scherbakov, A.S. Salasyuk, A.V. Akimov, X. Liu, M. Bombeck}},
  {\sc {D.R. Yakovlev, V.F. Sapega, J.K. Furdyna}}, and {\sc M.~Bayer},
\newblock {\em condmat} {\bf 1006.4133} (2010).

\bibitem{lemaitre08}
{\sc A.~Lema\^{\i}tre}, {\sc A.~Miard}, {\sc L.~Travers}, {\sc O.~Mauguin},
  {\sc L.~Largeau}, {\sc C.~Gourdon}, {\sc V.~Jeudy}, {\sc M.~Tran}, and {\sc
  J.~George},
\newblock {\em Appl. Phys. Lett.} {\bf 93}, 21123 (2008).

\bibitem{Cubukcu2010}
{\sc M.~Cubukcu}, {\sc H.~J. von Bardeleben}, {\sc K.~Khazen}, {\sc J.~L.
  Cantin}, {\sc O.~Mauguin}, {\sc L.~Largeau}, and {\sc A.~Lemaitre},
\newblock {\em Phys. Rev. B} {\bf 81}, 41202 (2010).

\bibitem{peronne06}
{\sc E.~P\'{e}ronne} and {\sc B.~Perrin},
\newblock {\em Ultrasonics} {\bf 44 Suppl 1}, e1203 (2006).

\bibitem{klatt09}
{\sc G.~Klatt}, {\sc M.~Nagel}, {\sc T.~Dekorsy}, and {\sc A.~Bartels},
\newblock {\em Electronics Letters} {\bf 45}, 310 (2009).

\bibitem{osc}
{Unpublished data from B. Perrin and E. Peronne} .

\bibitem{Glunk2009}
{\sc M.~Glunk}, {\sc J.~Daeubler}, {\sc L.~Dreher}, {\sc S.~Schwaiger}, {\sc
  W.~Schoch}, {\sc R.~Sauer}, {\sc W.~Limmer}, {\sc A.~Brandlmaier}, {\sc
  S.~T.~B. Goennenwein}, {\sc C.~Bihler}, and {\sc M.~S. Brandt},
\newblock {\em Phys. Rev. B} {\bf
  79}, 195206 (2009).

\bibitem{Koopmans2000}
{\sc B.~Koopmans}, {\sc {van Kampen M}}, {\sc J.~Kohlhepp}, and {\sc {de Jonge
  WJ}},
\newblock {\em Phys. Rev. Lett.} {\bf 85}, 844 (2000).

\bibitem{Kimel2005}
{\sc A.~Kimel}, {\sc G.~Astakhov}, {\sc A.~Kirilyuk}, {\sc G.~Schott}, {\sc
  G.~Karczewski}, {\sc W.~Ossau}, {\sc G.~Schmidt}, {\sc L.~Molenkamp}, and
  {\sc T.~Rasing},
\newblock {\em Phys. Rev. Lett.} {\bf 94}, 227203 (2005).

\bibitem{perrin96}
{\sc {B. Perrin, B. Bonello, J. C. Jeannet}} and {\sc E.~Romatet},
\newblock {\em Progress in Natural Science} {\bf S6}, 444 (1996).

\end{thebibliography}

\end{document}